\documentclass[10pt]{article}
\usepackage[utf8]{inputenc}
\usepackage{enumerate}
\usepackage[utf8]{inputenc}
\usepackage{amsmath}
\usepackage{amsfonts}
\usepackage{amssymb}
\usepackage{graphicx}
\usepackage{color}
\usepackage{mathptmx}
\usepackage{multicol}
\usepackage{float}
\RequirePackage[numbers,sort&compress]{natbib}
\RequirePackage[colorlinks=true,citecolor=blue,urlcolor=blue,linkcolor=blue]{hyperref}
\usepackage[left=2cm,right=2cm,top=2cm,bottom=2cm]{geometry}

\author{\textbf{Avijit Chowdhury}\footnote{Email: ac13ip001@iiserkol.ac.in}\hspace{8pt}\textbf{.}\hspace{8pt}\textbf{Narayan Banerjee}\footnote{Email: narayan@iiserkol.ac.in} \\ {\normalsize Department of Physical Sciences, 
Indian Institute of Science Education and Research Kolkata,
Mohanpur, WB 741 246, India}}
\title{{\LARGE\bf Superradiant stability of mutated Reissner-Nordstr\"{o}m black holes}}
\begin{document}
\date{}
\maketitle
\vspace{-0.5cm}
\section*{Abstract}
A scalar hair is used to generate a mutated black hole, which mimics an Einstein-Rosen bridge or a wormhole. The superradiant stability of the spacetime is studied under perturbation by an electrically charged massive scalar field. The system appears to be stable against superradiance irrespective of the mass of the test field.

\paragraph*{Keywords}Superradiance, Einstein-Rosen bridge, Scalar hair, Quasinormal modes

\section{Introduction}
The long-awaited detection of gravity waves came from violent phenomena of the merger of two compact objects\cite{ligo_PRL_2016,abott_PRL_2017, abott_GWTC_2018} and paved the way of a gravity wave astronomy. But radiations can emerge from black holes from more sedate perturbations to start with, some of which may finally result in other kinds of spectacular events like a black hole bomb\cite{press_Nature_1972}.

The phenomenon of a radiation picking up energy rather than disseminating it is known in physics for a long time\cite{ginzburg_1947}. The term ``superradiance'' to mark this process was coined by Dicke\cite{dicke_PR_1954}. For a systematic development of the subject, we refer to the thorough review by Brito, Cardoso and Pani\cite{brito_Springer_2015}. In the context of a black hole, such an enhancement of radiation arising out of the perturbation of a black hole by a test field is also quite a possibility. Initial investigations on black hole superradiance was summarized by Bekenstein and Schiffer\cite{bekenstein_PRD_1998}. More elaborate and recent developments can be found in~\cite{brito_Springer_2015}.

When a bosonic wave of frequency, less than a critical value, impinges upon a rotating black hole, the reflected wave gets amplified by extracting energy from the rotation of the black hole\cite{zeldovich_JETPL_1971,zeldovich_JETP_1972,vilenkin_PLB_1978}. This critical frequency is known as the superradiance frequency.

For Reissner-Nordstr\"{o}m (RN) black holes, such an amplification is possible for charged bosonic waves\cite{bekenstein_PRD_1973}. In this case, the superradiance frequency $\omega_c=q \Phi_H$, where $q$ is the electric charge of the impinging field, $\Phi_H=\frac{e}{r_+}$ is the electric potential at the event horizon of the black hole with charge $e$.

If this superradiantly amplified wave is reflected back into  the black hole by a potential barrier, the reflected wave is further amplified. Repeated reflections of the amplified wave ultimately result in superradiant instability of the black hole, popularly known as the ``black hole bomb''\cite{press_Nature_1972}. The mass ($\mu$) of the incident bosonic field normally needs to be more than the mode frequency ($\mu^2>\omega^2$) for producing a local minimum in the effective potential resulting in bound states of the bosonic wave and thus an instability\cite{damour_LNC_1976, cardoso_PRD_2004, cardoso_PRD_2004_err}.

Reissner-Nordstr\"{o}m black holes are known to be superradiantly stable against perturbation by massive charged scalar fields\cite{hod_PLB_2012,hod_PLB_2013,hod_PRD_2015,huang_EPJC_2016} in the entire parameter space and does not lead to a black hole bomb, whereas, rotating black holes (Kerr black holes) are not stable against perturbations by massive bosonic fields\cite{vilenkin_PLB_1978,press_Nature_1972, press_APJ_1973,zouros_AP_1979,detweiler_PRD_1980,dolan_PRD_2007,hod_PRD_2010,beyer_JMP_2011,myung_PRD_2011,hod_PLB_2012_1,brito_PRD_2013,dolan_PRD_2013,witek_PRD_2013, cardoso_GRG_2013,myung_PRD_2014,arderucio_ARXIV_2014}.

Kerr-Newman black holes, on the other hand, are superradiantly unstable against perturbation by massive charged scalar fields in a range of frequencies. The range is given by $\xi(\mu, q)<\omega<\omega_c<\mu$ or $\xi(\mu, q)<\omega<\mu<\omega_c$ (see Refs.\cite{huang_PRD_2016,huang_PRD_2018}). Here $\xi(\mu,q)$ is a quantity given as $\xi(\mu,q)\equiv \frac{e q}{4 M}+\sqrt{\frac{\mu^2}{2}+\frac{e^2 q^2}{16 M^2}}$ ; $\omega_c=m \Omega_H + q \Phi_H$ is the superradiance frequency with $\Omega_H$ and $\Phi_H$ being the horizon angular velocity and the horizon electric potential respectively and $m$ is the azimuthal harmonic index. $M$ and $e$ are the mass and electric charge of the black hole, whereas, $\mu$ and $q$ are the mass and electric charge of the perturbing scalar field. 

At the superradiance frequency, $\omega_c$, the imaginary part of $\omega$ vanishes and thus the modes do not grow or decay in time. These long-lived modes are dubbed as scalar clouds. Herdeiro and Radu\cite{herdeiro_PRL_2014} studied the superradiance of massive scalar fields in the background of a rotating black hole. They obtained a numerical solution describing a rotating black hole with a complex scalar field. Gravitational wave signals from ultra-light bosonic clouds around rotating black holes have been studied in detail in Ref.\cite{brito_PRL_2017, brito_PRD_2017, arvanitaki_PRD_2015, arvanitaki_PRD_2017}. 

Charged black holes in string theory are also superradiantly stable against massive charged scalar perturbation\cite{li_PRD_2013}. The recent work by Tokgoz\cite{tokgoz_ARXIV_2019} on dilaton black holes and references therein are also referred to in this context. Superradiance, in the context of scalar tensor gravity has been discussed by Cardoso, Carucci, Pani and Sotiriou\cite{cardoso_PRL_2013, cardoso_PRD_2013}.

Recently, Astorino\cite{astorino_PRD_2013} gave an asymptotically flat, static, spherically symmetric solution for the Einstein-Maxwell system conformally coupled to a scalar field. The metric is effectively similar to the RN black hole, except for the presence of a scalar charge $s$, which appears with power unity as an additional correction to the square of the electric charge $e$. It is important to note that this scalar hair is a primary hair as it retains its existence even if the electromagnetic field is switched off. The stress-energy tensor of the system is traceless, so the existence of such a ``primary hair'' is consistent with the result given in~\cite{narayan_pramana_2015}. For $s<-e^2$, the spacetime behaves as a ``mutated Reissner-Nordstr\"{o}m'' spacetime leading to an Einstein-Rosen bridge\cite{rosen_PR_1935}. The signature of the scalar hair, particularly when $s<-e^2$, in the quasinormal spectrum for various perturbing fields has been discussed in~\cite{chowdhury_EPJC_2018}.

The purpose of this work is the investigation of black hole  superradiance when the black hole has a primary scalar hair. This explores the existence of superradiance and the stability of the black hole against superradiance for a mutated RN black hole which mimics an Einstein-Rosen bridge. The idea is to check whether the superradiance condition and the bound state condition are satisfied simultaneously. We find that such a system is stable against superradiance. This is firmly established for a large mass of the test field, and is very strongly indicated for smaller masses as well.

The rest of the paper is organized as follows. In Sec.~\ref{section2}, starting from a charged black hole with scalar hair we  briefly introduce the mutated Reissner-Nordstr\"{o}m black hole and discuss the superradiance of massive charged test scalar field in this background. In Sec.~\ref{section3}, the possibility of the existence of bound states of the massive test field outside the event horizon is worked out. Sec.~\ref{section4} contains a summary and the definite conclusion that we arrive at.

\section{Black holes with a scalar hair and a mutated Reissner-Nordstr\"{o}m metric}\label{section2}
We consider the action of Einstein's gravity coupled to a Maxwell field $F^{\mu\nu}$ and conformally coupled to a self-interacting scalar field $\chi$,
\begin{equation}\label{eq_action}
I=\frac{1}{16\pi G}\int d^4 x\sqrt{-g}\left[ R-F_{\mu \nu}F^{\mu \nu}-8\pi G\left(\bigtriangledown_\mu \chi \bigtriangledown^\mu \chi+\frac{R}{6}\chi^2\right)\right].
\end{equation}
A static, spherically symmetric black hole of mass $M$ and electric charge $e$ endowed with a scalar hair $s$, in this theory is given by the metric\cite{astorino_PRD_2013},
 \begin{equation}\label{eq_metric}
ds^2=-f\left(r \right) dt^2+f\left(r \right)^{-1}dr^2+r^2 \left(d\theta^2 + \sin^2\theta d\phi^2 \right),
\end{equation}
where $f\left( r \right)=\left(1-\frac{2M}{r}+\frac{e^2+s}{r^2}\right)$ and $
\chi=\pm \sqrt{\frac{6}{8 \pi}}\sqrt{\frac{s}{s+e^2}}$.
The line element~(\ref{eq_metric}) will henceforth be referred to as the Reissner-Nordstr\"{o}m-scalar hair (shRN) black hole. 
The solution is written in gravitational units where $G=c=1$. \\
This solution is similar to the black hole solution given by Bekenstein\cite{bekenstein_AP_1974} for a conformally invariant scalar field. In the form (\ref{eq_metric}) as provided by Astorino\cite{astorino_PRD_2013}, the solution opens a new possibility. An Einstein-Rosen bridge or a wormhole, that connects two causally  disconnected spacetime\cite{rosen_PR_1935}, is given by the metric (\ref{eq_metric}) where $f = (1-\frac{2m}{r} - \frac{\epsilon^2}{r^2})$. In an RN solution, this last term, that determines the contribution of electric charge, always comes with a positive sign. So the Einstein-Rosen bridge is called a ``mutated'' RN solution. With $s$ having a large negative value (greater in magnitude that $e^2$), it is easy to see that the Astorino solution mimics this mutated RN solution.

The dynamics of a massive, charged test scalar field $\Psi$ (of mass $\mu$ and electric charge $q$) in the shRN background is governed by the Klein-Gordon equation,
\begin{equation}\label{eq_KG}
[\left(\nabla^\nu-iqA^\nu\right)\left(\nabla_\nu-iqA_\nu\right)-\mu^2]\Psi=0,
\end{equation}
where $A_\nu=-\delta^0_\nu e/r$ is the electromagnetic vector potential of the black hole.
The test field $\Psi$ can be decomposed as
\begin{equation}\label{eq_decom}
\Psi\left(t,r,\theta,\phi\right)=e^{-i\omega t}S_{lm}\left(\theta\right)R_{lm}\left(r\right)e^{i m \phi},
\end{equation}
where $\omega$ is the conserved frequency, $l$ is the spherical harmonic index and $m$ ($-l\leq m\leq l$) is the azimuthal harmonic index. Henceforth, the subscripts $l$ and $m$ will be dropped for brevity. Also, the test field $\Psi$ will involve a constant of intergration, which may be called the scalar charge of the test field. This is absorbed in the amplitude, which can be put equal to unity without any loss of generality. In general, $\omega$ is a complex quantity with the real part $\omega_{Re}$ corresponding to the actual frequency of the wave motion and the imaginary part $\omega_{Im}$ taking care of the damping. However, since any instability must set in through the real frequency modes (see Refs.~\cite{damour_LNC_1976,zouros_AP_1979,hartle_CMP_1974}), we only consider modes with $\mid\omega_{Im}\mid<<\omega_{Re}$.

With the decomposition~(\ref{eq_decom}), the Klein-Gordon equation~(\ref{eq_KG}) can be separated into a radial equation and an angular equation with the separation constant $K_l=l\left( l+1 \right)$. The radial Klein-Gordon equation is given by
\begin{equation} \label{eq_radial}
\Delta \frac{d}{dr}\left(\Delta \frac{dR}{dr}\right)+U R=0,
\end{equation}
where $\Delta=r^2 f\left(r\right)$ and $U=\left( \omega r^2-e q r \right)^2-\Delta \left[ \mu ^2 r^2+l \left( l+1 \right)\right]$.
The position of the inner and outer horizons are given by the roots of $\Delta$,
\begin{equation}
r_\mp=M \mp \sqrt{M^2-e^2-s}.
\end{equation}
In terms of the tortoise coordinate $r_{*}$ (defined by $dr_{*}=dr/f\left( r\right)$), mapping the semi infinite region $\left[r_{+},\infty\right)$ to $(-\infty,\infty)$, Eq.(\ref{eq_radial}) can be recast as
\begin{equation}\label{eq_trts}
\frac{d^2 \zeta}{d r_{*}^2}+W\left(\omega, r\right) \zeta=0,
\end{equation}
where $\zeta=r R \hspace{0.02\textwidth}\mbox{and}\hspace{0.02\textwidth} W\left(\omega, r\right)=\frac{U}{r^4}-\frac{\Delta}{r^3}\frac{d}{dr}\left(\frac{\Delta}{r^2}\right).$

For the scattering problem, with $\omega^2>\mu^2$, the physical boundary condition corresponds to an incident wave of amplitude $\mathcal{I}$ from infinity giving rise to a reflected wave of amplitude $\mathcal{R}$ near infinity and a transmitted wave of amplitude $\mathcal{T}$ at the horizon,
\begin{equation}\label{eq_BC_superrad}
\zeta \sim 
\begin{cases}
\mathcal{I}e^{-i\sqrt{\omega^2-\mu^2} r_*}+\mathcal{R}e^{i\sqrt{\omega^2-\mu^2} r_*} &\mbox{as} \quad r_*\rightarrow \infty\\
 \mathcal{T}e^{-i \left(\omega-\frac{e q}{r_{+}} \right) r_*}&\mbox{as} \quad  r_* \rightarrow -\infty. 
\end{cases}
\end{equation}
The invariance of the field equation under the transformation $t\rightarrow-t$ and $\omega\rightarrow-\omega$, leads to another linearly independent solution $\zeta^{*}$ which satisfies the complex conjugate boundary conditions. The Wronskian of the two solutions will be independent of $r_{*}$. Hence, the Wronskian evaluated at the horizon, $W_h= 2i\left(\omega-\frac{e q}{r_{+}} \right)|\mathcal{T}|^2$, must be equal to that evaluated near spatial infinity, $W_\infty=-2i\sqrt{\omega^2-\mu^2}\left(|\mathcal{R}|^2-|\mathcal{I}|^2\right)$. This yields,
\begin{equation}
|\mathcal{R}|^2=|\mathcal{I}|^2-\frac{\omega-\frac{e q}{r_+}}{\sqrt{\omega^2-\mu^2}}|\mathcal{T}|^2.
\end{equation}
The reflected wave is thus superradiantly amplified $\left( |\mathcal{R}|^2>|\mathcal{I}|^2 \right)$,  provided
\begin{equation}\label{eq_sprrdnc}
\omega<\frac{e q}{r_+}.
\end{equation}
However, when $\omega^2<\mu^2$, Eq.(\ref{eq_trts}) results in bound states of the scalar field characterised by exponentially decaying modes near spatial infinity, 
\begin{equation}\label{eq_bndst1}
\zeta \sim 
 e^{-\sqrt{\mu^2 -\omega^2} r_*}\mbox{\hspace*{2mm} as \hspace*{2mm}} r_*\rightarrow \infty.
\end{equation}

As mentioned earlier, we are particularly interested in studying the superradiant stability of the shRN black hole with large negative values of the scalar charge,
\begin{equation}\label{eq_mtdcondtn}
s=-|s|\hspace{0.5cm} \mbox{and} \hspace{0.5cm} |s |> e^2.
\end{equation}
The assumption~(\ref{eq_mtdcondtn}) results in  $r_-<0$, which is unphysical and hence the mutated RN spacetime is characterised by only one event horizon at $r=r_{+}$.

\section{Existence of bound states}\label{section3}
In this section, we investigate the existence of bound states of the scalar field in the superradiant regime. We present the radial Klein-Gordon Eq.(\ref{eq_radial}) in a Schr\"{o}dinger like form by defining a new radial function $\psi (r) =\sqrt{\Delta} R$, as
\begin{equation}
\frac{d^2 \psi}{dr^2}+\left(\omega^2-V \right)\psi=0,
\end{equation}
where
\begin{equation}\label{eq_pot}
V=\omega^2-\frac{U+M^2-e^2-s}{\Delta^2}
\end{equation}
and analyse the nature of the effective potential $V$ to check for the existence of a potential well outside the event horizon supporting meta-stable bound states in the superradiant regime. In the asymptotic limit,
\begin{equation}\label{eq_potlimit1}
V\left(r\rightarrow \infty\right)\rightarrow
 \mu ^2+\frac{2~a\left(\omega\right)}{r}+\mathcal{O}\left(\frac{1}{r^2}\right),
\end{equation}
where
\begin{equation}\label{eq_a}
a\left(\omega\right)=e q \omega +\mu ^2 M-2 M \omega ^2
\end{equation}
represents a convex parabola. The sign of $a\left(\omega\right)$ at the edges completely determines the asymptotic behaviour of the effective potential.

The superradiance condition $\left( \omega<e q/r_{+} \right)$ and the bound state condition $\left( \omega^2<\mu^2 \right)$ can be combined to yield
\begin{equation}\label{eq_omegalimit}
0\leq\omega<\mbox{min}\left\lbrace\tfrac{e q}{r_+},\mu \right\rbrace.
\end{equation}
 The left hand bound is required as $\omega$ is the frequency. The behaviour of $a\left(\omega\right)$ near the boundaries of~(\ref{eq_omegalimit}) can be summarized as
\begin{equation}
a\left(\omega\right)\rightarrow
\begin{cases}
 \mu^2 M >0 & \mbox{as} \quad \omega\rightarrow 0 \\
\mu\left(e q-M\mu \right)>\mu M\left(\frac{e q}{r_+}-\mu \right) >0 & \mbox{as} \quad \omega\rightarrow\mu \quad \mbox{for} \quad \mu<\tfrac{e q}{r_+}\\
M\mu^2+\frac{e^2 q^2}{r_+}\left(1-\frac{2 M}{r_+}\right)>0 & \mbox{as} \quad \omega\rightarrow\tfrac{e q}{r_+} \quad \mbox{for} \quad \tfrac{e q}{r_+}<\mu.
\end{cases}
\end{equation}
Thus,
\begin{equation}\label{eq_a>0}
a\left(\omega\right)>0
\end{equation}
in the entire range~(\ref{eq_omegalimit}). Further,
\begin{equation}\label{eq_potlimit2}
V\rightarrow
\begin{cases}
-\infty & \mbox{ as \hspace*{2mm}} r\rightarrow r_{+} \\
 \omega ^2-\frac{l(l+1)+1}{|s |-e^2}-\frac{M^2}{\left(|s |-e^2\right)^2}
 & \mbox{ as \hspace*{2mm}} r\rightarrow 0 \\
-\infty & \mbox{ as \hspace*{2mm}} r\rightarrow r_{-} .
\end{cases}
\end{equation}
From Eq.\eqref{eq_potlimit1} and  \eqref{eq_a>0}, we note that $V(r\rightarrow \infty)$ is positive definite and in light  of Eq.\eqref{eq_potlimit2} (namely, $V(r\rightarrow r_+)\rightarrow -\infty$), we infer that $V$ has at least one maximum outside the horizon $\left(r_{+}<r<\infty\right)$. Eq.\eqref{eq_potlimit2} also indicates that $V$ has another maximum in the unphysical region, $r_{-}<r<r_{+}$.

The derivative of the effective potential is given by 
\begin{equation}
\begin{split}
V^{'}=&-\tfrac{2}{\Delta^3}\left[a\left(\omega\right)r^4+\left[-2M^2 \mu^2-e^2 q^2+\left(|s|-e^2\right)\left(\mu^2 -2 \omega^2 \right)+2 M e q \omega+l\left(l+1 \right)\right]r^3 \right.\\
&\left.+3\left[-M \mu^2 \left(|s|-e^2 \right)+eq\omega\left(|s|-e^2  \right)-M l\left(l+1\right)\right]r^2 +\left[l \left(l+1\right)\left(e^2+2 M^2-|s|\right)-2 \left(M^2+|s|-e^2 \right) \right.\right.\\
& \left.\left. -e^2 q^2 \left(|s|-e^2\right)-\mu ^2 \left(|s|-e^2\right)^2 \right]r+2 M \left(M^2+|s|-e^2 \right) +l (l+1) M \left(|s|-e^2\right)\right],
\end{split}
\end{equation}
which in the asymptotic limit reduces to $V^{'}(r\rightarrow\infty)\rightarrow 0^-$, suggesting the absence of any potential well as $r\rightarrow\infty$.

If we define $z=r-r_{-}\left(=r+|r_- |\right)$, and write $V^{'}$ as a function of $z$, one has
\begin{equation}\label{eq_V'(z)}
V^{'}\left(z \right)=-\frac{2}{\Delta^3}\left(a z^4+bz^3+c z^2+d z+g \right),
\end{equation}
where
\begin{align}
\begin{split}
b={ }& -\mu ^2 \left(e^2+2 M^2+4 M |r_{-}|-|s|\right)+2 \omega ^2 \left(e^2+4 M |r_{-}|-|s|\right) -e^2 q^2+2 e q \omega  (M-2 |r_{-}|)+l (l+1),
\end{split}\\
\begin{split}
c={ }& 3|r_{-}|^3\left( \tfrac{e q}{|r_{-}|}+\omega \right) \left( \tfrac{e q}{|r_{-}|}+2\omega \right)-3\mu ^2 |r_{-}|^2 \left(r_{+}-M \right)-3l \left(l+1 \right) \left(r_{+}-M \right),\label{eq_c}
\end{split}\\
\begin{split}
d={ }& 2 \mu ^2 |r_{-}|^2 \left(M^2+|s|-e^2\right)-e^2 q^2 |r_{-}|\left(r_{+}+3 |r_{-}|\right)-2 e q |r_{-}|^2 \omega  \left(3 r_{+}+2 |r_{-}|-3 M \right)\\
& -2 |r_{-}|^3 \omega ^2 \left(3 r_{+}-4 M \right)+2 \left(l \left(l+1 \right)-1\right) \left(M^2+|s|-e^2 \right),
\end{split}\\
g={ }& 2 |r_{-}|^4 (r_{+}-M) \left(\tfrac{e q}{|r_{-}|}+\omega \right)^2+2 (r_{+}-M)^3.\label{eq_e}
\end{align}

If ${z_1,z_2,z_3,z_4}$ are the roots of the equation $V^{'}(z)=0$, then Vieta's formulas \cite{barnard_child_1959} give
\begin{eqnarray}
&z_1+z_2+z_3+z_4=-\frac{b}{a}\label{eq_z1+z2+z3+z4},\\
&z_1z_2+z_1z_3+z_1z_4+z_2z_3+z_2z_4+z_3z_4=\frac{c}{a}\label{eq_z1z2+},\\
&z_1z_2z_3+z_1z_2z_4+z_1z_3z_4+z_2z_3z_4=-\frac{d}{a}\label{eq_z1z2z3+},\\
\mbox{and} &z_1z_2z_3z_4=\frac{g}{a}.\label{eq_z1z2z3z4}
\end{eqnarray}
The existence of at least one maximum of the effective potential outside the horizon guarantees that $V^{'}(z)$ has at least one positive root $\left(z_1\mbox{, say}\right)$. Similarly, the potential maximum in the unphysical region $r_{-}<r<r_{+}$ suggests another positive root of $V^{'}(z)$ $\left(z_2\mbox{, say}\right)$ with
\begin{equation}\label{eq_z1z2>0}
z_1>z_2>0.
\end{equation}
Eq.(\ref{eq_e}) shows 
\begin{equation}
g>0
\end{equation}
which together with Eqs.(\ref{eq_a>0}) and~(\ref{eq_z1z2z3z4}) implies
\begin{equation}\label{eq_z1z2z3z4>0}
z_1z_2z_3z_4>0.
\end{equation}
Combining Eqs.(\ref{eq_z1z2>0}) and~(\ref{eq_z1z2z3z4>0}) one deduces that $z_3$ and $z_4$ must be of the same sign,\begin{equation}\label{eq_z3z4>0}
z_3z_4>0.
\end{equation}

For a potential well to exist outside the horizon, beyond $z_1$, the roots $z_3$ and $z_4$ must be real and positive with
\begin{equation}\label{eq_z3,z4>z1}
z_3,z_4>z_1.
\end{equation}
Eq.(\ref{eq_z3,z4>z1}) in conjunction with Eqs.(\ref{eq_a>0}),(\ref{eq_z1+z2+z3+z4}),(\ref{eq_z1z2+}),(\ref{eq_z1z2z3+}),(\ref{eq_z1z2>0}) and~(\ref{eq_z3z4>0}) implies
\begin{equation}\label{c>0b<0}
b\leq 0, \quad c\geq 0 \quad \mbox{and} \quad d\leq 0.
\end{equation}
If we drop the last term in the expression of $c$ in Eq.(\ref{eq_c}) (which is at most zero) and define
\begin{equation}
\tilde{c}=3|r_{-}|^3\left( \tfrac{e q}{|r_{-}|}+\omega \right) \left( \tfrac{e q}{|r_{-}|}+2\omega \right)-3\mu ^2 |r_{-}|^2 \left(r_{+}-M \right),
\end{equation}
then $\tilde{c}$ represents a concave parabola which crosses the $\omega$-axis at 
\begin{eqnarray}
&\omega_1=\frac{-3eq-\sqrt{e^2q^2+4|r_{-}|\left(r_{+}+|r_{-}|\right) \mu^2}}{4|r_{-}|} < 0\\
\mbox{and} &\omega_2=\frac{-3eq+\sqrt{e^2q^2+4|r_{-}|\left(r_{+}+|r_{-}|\right) \mu^2}}{4|r_{-}|}.
\end{eqnarray}

As $\omega_1$ is a negative definite quantity and we are looking for the possibility of an enhanced radiation, we shall work with $\omega_2$. We separate the parameter space of the massive charged test scalar field in three regions based on the mass of the test field.
\begin{enumerate}[{\bf \mbox{Region} I:}]
\item \label{regionI} \begin{equation}
\mu\geq \mu_1 \left(>\frac{e q}{r_{+}} \right),
\label{eq_massivelimit}
\end{equation}
where
\begin{equation}
\mu_1=\frac{e q}{r_{+}}\sqrt{\frac{2\left(r_{+} + 2|r_{-} |\right)}{|r_{-} |}}.
\end{equation}
For such values of the field mass, one finds
\begin{equation}
\omega_2\geq \tfrac{e q}{r_{+}}>0
\end{equation}
and since the concave parabola $\tilde{c}$ crosses the $\omega$-axis at $\omega_1(<0)$ and $\omega_2$, for any $\omega$ in the region $0\leq\omega<\frac{e q}{r_+}$, one has
\begin{equation}\label{eq_c<0}
c\leq \tilde{c}<0.
\end{equation}
The inequality~(\ref{eq_c<0}) contradicts condition (\ref{c>0b<0}). Hence, there is no potential well in the physical region $r>r_{+}$. Thus for $\mu$ satisfying the condition~(\ref{eq_massivelimit}), there is no bound state and so the mutated black hole is stable against superradiance.

\item \label{regionII} 

This region is defined as
\begin{equation}\label{eq_w2=0}
\left(\frac{e q}{r_+} <\right)\mu_2<\mu<\mu_1,
\end{equation}
where
\begin{equation}
\mu_2=e q\sqrt{\frac{2}{|r_{-} |\left(r_{+}+|r_{-} |\right)}}.
\end{equation}
Now, $\omega_2$ lies in the range
\begin{equation}
0<\omega_2<\tfrac{e q}{r_{+}},
\end{equation}
and there exists an $\omega$, in the range $0\leq\omega<\omega_2$, for which $\tilde{c}< 0$, in contradiction with~(\ref{c>0b<0}) and as before, the spacetime is superradiantly stable.

\item \label{regionIII} 

\begin{equation}\label{eq_w2<0}
0<\mu\leq \mu_2,
\end{equation}
For $\mu$ in this range, $\omega_2\leq 0$ resulting in $\tilde{c}>0$ for any $\omega>0$. The coefficients $b$ and $d$ can also be shown to be negative for $l=0$ in this range. To further ascertain the absence of potential well in this region, we analyse the signature of the discriminant of Eq.\eqref{eq_V'(z)}.
\end{enumerate}

It is well known that a negative discriminant of quartic polynomial implies the existence of two real and two complex roots (see Ref.~\cite{rees_JSTOR_1922}). Table~\ref{table1} summarizes the signature of the discriminant $(D)$ of $V^{'}(z)$ and the nature of $z_3$ and $z_4$ at some values of $\mu$ for $l=0$ at the boundaries of the superradiant regime. The expression for $D$ is,
\begin{align}\label{eq_dscrmnt}
D& =-2 a b d \left(96 a g^2+40 c^2 g-9 c d^2\right)+b^2 \left(144 a c g^2-6 a d^2 g-4 c^3 g+c^2 d^2\right)\nonumber \\
& \quad +a \left(-128 a c^2 g^2+144 a c d^2 g+a \left(256 a g^3-27 d^4\right)+16 c^4 g-4 c^3 d^2\right)-27 b^4 g^2+b^3 \left(18 c d g-4 d^3\right).
\end{align}

\begin{table}[htbp!]
\centering
\caption{Table showing the signature of the discriminant $D$ in the superradiant regime at some discrete values of $\mu$.}
\begin{tabular}{|c|c|c|c|c|c|}
\hline
$\mu\rightarrow$    & \multicolumn{2}{c|}{\begin{tabular}[c]{@{}c@{}}$\mu_1$\end{tabular}} & \multicolumn{2}{c|}{$\mu_2$} & 0 \\ \hline
$\omega\rightarrow$ & 0 & $\frac{e q}{r_{+}}$ & 0 & $\frac{e q}{r_{+}}$ & 0  \\ \hline
Sign$(D)$ & $<0$ (for $|s|-e^2\leq 4.414 M^2$) & $<0$ & $<0$ & $<0$ & $<0$ \\ \hline
Nature of $z_3,z_4$ & Complex  (for $|s|-e^2\leq 4.414 M^2$); Real, negative otherwise & Complex & Complex & Complex & Complex \\
\hline
\end{tabular}
\label{table1}
\end{table} 
We observe from table~\ref{table1} that at the boundaries of~(\ref{eq_w2=0}) and~(\ref{eq_w2<0}),  $z_3$ and $z_4$ are complex in the superradiant regime. Thus, there exists no potential well and hence the spacetime is expected to be superradiantly stable, irrespective of the mass of the test field.
 \begin{figure}[H]
\centering
\includegraphics[width=0.75\textwidth]{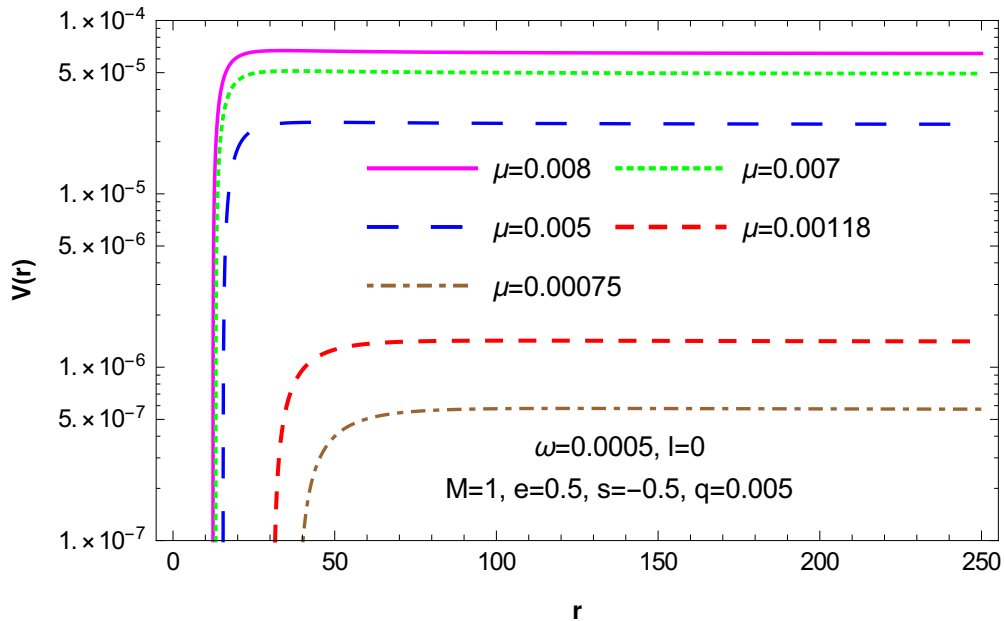}
\caption{Variation of the effective potential $V(r)$ with $r$ in logarithmic scale for different values of the field mass $\mu$.}
\label{fig:1}
\end{figure}
Fig.\ref{fig:1} shows the potential profile of a mutated RN black hole of unit mass with electric charge $e=0.5$ and scalar charge $s=-0.5$ for different values of the field mass for the $l=0$ mode in logarithmic scale. The event horizon of the black hole is at $r_+=2.118$. The electric charge of the test field $q=0.005$ which yields a superradiance frequency $e q/r_+=0.00118$. The boundaries of region~\ref{regionI} and region~\ref{regionII} are at $\mu_1=0.00745$ and $\mu_2=0.00688$. We observe that no potential well exist outside the event horizon when $\mu$ lies in region~\ref{regionI} ($\mu>\mu_1$). This observation also holds when $\mu$ lies in region~\ref{regionII} ($\mu_2<\mu<\mu_1$) with $\omega_2(=0.00024)<\omega(=0.0005)$ and even for $\mu<\mu_2$.

\section{Summary and Discussion}\label{section4}
In the present work, we investigated the possibility of the existence of superradiance and the superradiant stability of a charged spherically symmetric black hole with scalar hair, particularly for $s<-e^2$, representing a mutated RN spacetime. The quasinormal modes of the mutated RN spacetime against perturbation by massive (and massless) charged (and uncharged) scalar fields have been rigorously studied in Ref.\cite{chowdhury_EPJC_2018}. Those quasinormal frequencies were obtained with negative imaginary part implying that the perturbations decay in time and the corresponding modes are stable.

In this work, we considered modes with $Re(\omega)\gg |Im(\omega)|$ and evaluated the superradiance condition for the mutated RN spacetime (see~(\ref{eq_sprrdnc})). The superradiance frequency is found to be consistent with that of a standard RN black hole\cite{bekenstein_PRD_1973,hod_PLB_2012,hod_PLB_2013,hod_PRD_2015,huang_EPJC_2016}.

We observe that in the superradiant regime, if the mass of the test field lies in region~\ref{regionI}, there is no potential well outside the horizon and hence no bound state resonance of the massive scalar fields. The spacetime is thus superradiantly stable in the  range~(\ref{eq_sprrdnc}). If the mass of the test scalar lies in region~\ref{regionII}, the stability can be conclusively proved in the range of frequencies, $0\leq\omega<\omega_2$ with $0<\omega_2<e q/r_+$.

In region~\ref{regionIII} where $\mu<\mu_2$, Descartes' rule of signs eliminates the possibility of negative real roots of $V^{'}(z)$, so we investigated the signature of the discriminant ($D$) of $V^{'}(z)$ for the lowest angular momentum state (see~(\ref{eq_dscrmnt})). We note that as $\mu\rightarrow \mu_2$, $D<0$ at both ends of the superradiant regime~(\ref{eq_sprrdnc}) suggesting the existence of two complex roots ($z_3,z_4$) and thus confirms the superradiant stability of the spacetime. We observe that for vanishingly small field masses in the superradiant regime, $D<0$, which is expected, as for very small values of the field mass the potential well will almost cease to exist and the spacetime will be superradiantly stable. Further, as $\mu\rightarrow\mu_1$ and $\omega\rightarrow 0$, $z_3,z_4$ are complex for $|s|-e^2\leq 4.414M^2$ and real negative otherwise. In either situation, there will be no potential well outside the horizon and the spacetime will be superradiantly stable.

We conclude that the mutated RN black hole is superradiantly stable. For a large mass of the test field, this stability is proved. As with smaller masses, the strength of the well should decrease, the stability is expected to be ensured. However, such a comprehensive proof could not be provided for smaller masses. But the signature analysis of the discriminant quite strongly indicates a stable superradiance rather than a black hole bomb, for smaller masses as well. The potential profile shown in Fig.\ref{fig:1} for different values of the field mass also supports this conclusion.

Albeit a mutated RN spacetime is qualitatively different from an RN black hole, the results in connection with superradiance is very much similar in the two cases\cite{hod_PLB_2012,hod_PLB_2013,hod_PRD_2015,huang_EPJC_2016}. Calculations for the case,  $0<s+e^2<M^2$ are omitted, as it would effectively resemble an RN black hole (see metric\eqref{eq_metric}) and no characteristic difference from the RN black hole is expected.
%

\end{document}